\documentclass[aps,twocolumn,showpacs]{revtex4}
\usepackage{amsfonts,amsmath,amssymb}
\usepackage{graphicx}

\begin{document}

\title{Quaternionic and Hyper-K\"{a}hler metrics \\
from Generalized Sigma models}
\author{V.I. Afonso$^{1}$, D. Bazeia$^{1,2}$, D.J. Cirilo-Lombardo$^{3,4}$}
\affiliation{$^1$Unidade Acad\^{e}mica de F\'{\i}sica, Universidade Federal de Campina
Grande, PB, Brasil}
\affiliation{$^2$Departamento de F\'{\i}sica, Universidade Federal da Para\'{\i}ba, PB,
Brasil}
\affiliation{$^{3}$International Institute of Physics, Natal, RN, Brasil and $^{4}$%
Bogoliubov Laboratory of Theoretical Physics, Joint Institute for Nuclear
Research, Dubna, Russian Federation}

\begin{abstract}
The problem of finding new metrics of interest, in the context of SUGRA, is
reduced to two stages: first, solving a generalized BPS sigma model with
full quaternionic structure proposed by the authors and, second,
constructing the hyper-K\"{a}hler metric, or suitable deformations of this
condition, taking advantage of the correspondence between the quaternionic
left-regular potential and the hyper-K\"{a}hler metric of the target space.
As illustration, new solutions are obtained using generalized Q-sigma model
for Wess-Zumino type superpotentials. Explicit solutions analog to the
Berger's sphere and Abraham-Townsend type are given and generalizations of
4-dimensional quaternionic metrics, product of complex ones, are shown and
discussed.
\end{abstract}

\maketitle




\section{Introduction}

Several attempts have been proposed in order to find new metric structures
for the target space in supersymmetric models, in particular the $\mathcal{N}%
=(4,0)$ and the $\mathcal{N}=(4,4)$ cases. Since each supersymmetry, beyond
the first, requires the existence of complex or quaternionic structure,
these attempts are expected to lead to interesting new hypercomplex
geometries in the context of SUGRA and, consequently, in type IIA and type
IIB superstring theories. Considerable efforts and beautiful methods and
prescriptions have been developed in that direction: from the bosonic
approach, monopole solutions in S$^{3}$ sub manifolds \cite{papad1, CTV} and
the method of the calibrations \cite{papad2} (and refs. therein); from the
supersymmetric side, the harmonic superspace method \cite{ivaval, ivadel}.
However, tentatives to connect the truly BPS solutions of the nonlinear
sigma model under consideration, with the corresponding geometries (metrics)
showing the expected hyper-K\"{a}hler and quaternionic properties, remain
lacking.

In this work, our main goal is to attack this problem by reducing it to two
independent stages consisting in: $(i)$ solving a generalized BPS Sigma
model with full quaternionic structure and $(ii)$, with this information,
constructing the hyper-K\"{a}hler metric using the correspondence between
the quaternionic left-regular potential and the hyper-K\"{a}hler metric in
the target space. This correspondence, valid for hyper-K\"{a}hler manifolds
and quaternionic ones with compact substructure (also smooth departures of
the hyper-K\"{a}hler condition are allowed), is based on the existence of
certain geometrical mappings that, at the classical level, allow to
establish the equivalence between \textit{a model with a potential in flat
spacetime} and \textit{a free model with a suitable metric in the target
space}.

The first step of our strategy consists in finding BPS solutions for a
Generalized Quaternionic Lagrangian ($GQL$), as introduced in ref. \cite%
{tobepub}, which presents the important property that both the base and the
target spaces live in $\mathbb{H}$ (see subsection below). As we will show
in the following, the proposed $GQL$ has a standard form with a potential
depending on scalar (quaternionic) fields. Then, with solutions for this
model at hand, we will be able to establish a correspondence with the even
sector ($B_{0}$) of a supergravity theory. The link is realized by a direct
mapping between the action of the free sigma model with metric $g_{\mu \nu }$
and the $GQL$ action in a Minkowski space with a potential $V(q)$.

It is worth mentioning that the considered model allows, for certain choices
of the coset, cohomogeneity one metric solutions, a type well studied in the
context of Spin(7) manifolds -- see, for instance \cite{CGLP}.

Due to their simplicity and the clear importance they have in the context of
supersymmetric nonlinear sigma models, through this letter we will focus our
study on Wess-Zumino type potentials. This will allow us to present the
analysis by showing specific explicit solutions, which will put in evidence
the underling quaternionic structures behind them.


\section{Generalized Quaternionic Action}

In all the considered cases, our model is defined on a quaternionic
spacetime (base space) and with \emph{a quaternionic space of scalar fields
as target}. This is the correct and clear geometrical definition. Another
possible terminology (not strictly mathematically accurate) is \emph{%
worldvolume} and \emph{target space}, in referring to the domain and range
manifolds, respectively, of the sigma models \cite{Sezgin}

Because of the obvious group relation $\mathbb{R\subset C\subset H}$, we can
take a standard real 4-dimensional spacetime, or a 2-dimensional complex or
a 1-dimensional quaternionic manifold as our base space to the quaternionic
(target) space of the (four) scalar fields.

Consider a system of four scalar fields governed by the Generalized
Quaternionic Lagrangian density ($GQL$) of the form 
\begin{equation}
\mathcal{L}=\frac{1}{2}\overline{\Pi q}\,\Pi q-\frac{1}{2}\left\vert
W^{\prime}(q) \right\vert ^{2}.
\end{equation}
The \textit{Cauchy-Fueter operator} $\Pi$ is defined by 
\begin{equation}
\Pi\equiv\widehat{i^{0}}\partial_{0}-\widehat{i^{1}}\partial_{1}-\widehat{%
i^{2}}\partial_{2}-\widehat{i^{3}}\partial_{3},
\end{equation}
where $\widehat{i^{0}}=\mathbb{I}$ and $\widehat{i^{i}}$ ($i=1,2,3$) obey
the standard quaternionic algebra, and $\partial_{0}\equiv{\partial }/{%
\partial x^{0}}$ and $\partial_{i}\equiv{\partial}/{\partial x^{i}}$.

Throughout this work we shall adopt Einstein's convention of indices
summation, with the Latin indices $i$ and $j$ running from $1$ to $3$,
unless otherwise stated. As usual, $Sc$ and $Vec$ will denote the scalar and
vector parts of the corresponding quaternionic expression. In particular,
whenever convenient, we will use the notation $\Pi_{0}\equiv Sc\Pi$ and ${%
\mathbf{\Pi}}\equiv Vec\,\Pi$. We also define $q_{i}^{2}\equiv
q_{1}^{2}+q_{2}^{2}+q_{3}^{2}$, and $W_{q_{0}}\equiv\Pi_{q_{0}}W(q)$ and $%
W_{q_{i}}\equiv{\mathbf{\Pi}}_{q}W(q)$, where the Cauchy-Fueter operator
acting on the target space is given by 
\begin{equation}
\Pi_{q}\equiv\widehat{i^{0}}\partial_{q_{0}}-\widehat{i^{1}}\partial_{q_{1}}-%
\widehat{i^{2}}\partial_{q_{2}}-\widehat{i^{3}}\partial_{q_{3}}.
\end{equation}
Note that these equations for $W$ present both the scalar and vector parts.


\subsection{Wess-Zumino model}

Following the lines of the complex field case treated in ref. \cite{complex}%
, in the present work we will specialize our Generalized Q-sigma Model to
the case of a Wess-Zumino ($WZ$) type superpotential \cite{WZ} of the form 
\begin{equation}
W^{\prime }(q)=n-q^{N}=n-(q_{0}+\widehat{i^{i}}\,q_{i})^{N},  \label{WZpot}
\end{equation}%
where $N\in \mathbb{Z}$, in principle $n\in \mathbb{H}$ but through this
letter we will take $n\in \mathbb{C}$ or its subgroups, and the prime
indicates derivative with respect to the argument of the considered
function. This standard choice is simply motivated by the fact that the $WZ$
superpotential is the basic prototype for any analysis involving
hypercomplex quaternionic structures in several areas of the modern
theoretical physics.

Thus, the first order equation $\Pi\, q=\overline{W^{\prime}(q)}$ for our $%
WZ $ potential (\ref{WZpot}) reads 
\begin{equation}  \label{1stord}
\frac{dq}{dx} =n-\overline{q}^{N}=n-(q_{0}-\widehat{i^{i}}q_{i})^{N}
\end{equation}
This expression, with $x$ identified below, arises from the relation between
the left regular superpotential $W(q)$ and the BPS conditions as \textit{%
quaternionic configurations with left regular superpotentials minimize the
energy of the system to the BPS bound.} (see Appendix \ref{Appndx}).

The corresponding vacuum (minima) manifold is described by the set of the $N$%
-roots of the unity in the field of the quaternions, \emph{i.e.} $S^{2}$
spheres, 
\begin{equation}
v_{N}^{k}=e^{a_{N}^{k}}=\exp\left( a2\pi\frac{k-1}{N}\right) \qquad k=1,
\ldots, N,
\end{equation}
being $a$ a pure quaternion of unitary norm.

In the following we will present some solutions (orbits) for the $WZ$ model
above, in the simplest cases of $N=1$ and $N=2$. As a first approach to the
problem, we will consider an ordinary (commutative) base space as spacetime
equivalent. Then, we will focus on the case of a quaternionic
(noncommutative) base space as spacetime equivalent.


\subsection{Commutative spacetime equivalent solutions}

The realization of the commutative base space as spacetime equivalent is
achieved by making the identification 
\begin{equation}
x\rightarrow\widehat{i}_{0}X^{0}\quad(i.e. \; x\in\mathbb{R}).
\end{equation}


\subsubsection{Case $N=1$ (commutative)}

For $N=1$, equation (\ref{1stord}) reads 
\begin{equation}
\frac{dq}{dx}=n-\overline{q}=n-q_{0}+\widehat{i^{1}}q_{1}+\widehat{i^{2}}%
q_{2}+\widehat{i^{3}}q_{3}.
\end{equation}
Splitting up this equation into its $Sc$ and $Vec$ parts, we have 
\begin{align}
\frac{dq_{0}}{dx} & =n-q_{0} \\
\frac{dq_{i}}{dx} & =q_{i}
\end{align}
This system admits the direct solution 
\begin{equation}
q(x)=n-C_{0}\,e^{-x}+\widehat{i^{i}}\,C_{i}\,e^{x},  \label{solN1comm}
\end{equation}
where $C_{0}$ and $C_{i} \;(i=1,2,3)$ are integration constants.


\subsubsection{Case $N=2$ (commutative)}

For $N=2$, the first order equation takes the form 
\begin{equation}  \label{1stordN2}
\frac{dq}{dx} =n-(q_{0}^{2}-q_{i}^{2}-2\,\widehat{i^{i}}\,q_{i} q_{0}),\quad
n\in\mathbb{Z}.
\end{equation}
Given the condition on the quaternionic phase and the $Vec(q)$, this case
presents two minima in the field space, located at $Sc(q)=\pm1$ (analogously
to the complex field case of \cite{complex}).

Breaking equation (\ref{1stordN2}) into its $Sc$ and $Vec$ parts, we obtain
the system of equations 
\begin{eqnarray}
\frac{dq_{0}}{dx}&=&n-q_{0}^{2}+q_{i}^{2} \\
\frac{dq_{i}}{dx}&=&2q_{i}q_{0}.
\end{eqnarray}

Let us now propose a generic equation for the curves connecting the two
minima of the potential (trial orbit method \cite{trial}), 
\begin{equation}
A\,q_{0}^{2}+B^{i}q_{i}^{2}=C \quad\rightarrow\quad
q_{0}^{2}+\beta^{i}q_{i}^{2}=1
\end{equation}
where the reduction from five ($A, B_{i}, C$) to three ($\beta_{i}$)
parameters is due the $N=2$ condition. Differentiating the orbit equation
with respect to $x$, and using the first order derivatives of the system
above we obtain%
\begin{equation}  \label{orbitalpha}
q_{0}^{2}+\alpha^{i}q_{i}^{2}=n,\qquad\alpha_{i}=-(1+2\beta_{i})\quad
\forall i,
\end{equation}
from which it is easily seen by simple comparison that acceptable
hypercurves must fulfill the ellipticity conditions $\forall
\beta_{i}\leqq-1/2$ ($\alpha_{i}\geq 0$) and $n=1$. Thus, equation (\ref%
{orbitalpha}) reduces to 
\begin{equation}
q_{0}^{2}+\alpha^{i}q_{i}^{2} =1,\qquad\alpha_{i}\geq0 \quad\forall i.
\end{equation}
Imposing the orbit condition on the first order equations leads to, 
\begin{eqnarray}
\frac{dq_{0}}{dx}&=& q_{i}^{2} +\alpha^i q_{i}^{2} \\
\frac{dq_{i}}{dx}&=&\pm2q_{i}\sqrt{1-\alpha^{i}q_{i}^{2}}
\end{eqnarray}
Leaving aside the trivial case $\alpha_i=0 \;\forall i$(which leads to $%
q_0=\pm 1, q_i=0\;\forall i$), a solution to the system above presents
components of the form 
\begin{equation}  \label{solN2}
q_{0}(x)=\tanh(2x+c),\quad q_{i}(x)=\frac{1}{\sqrt{3\alpha_i}} \,\mbox{sech}%
(2x+c),
\end{equation}
where $a_i$, and $c$ are integration constants. Requiring consistency with
the orbit under consideration results in the restriction $%
1/\alpha_1+1/\alpha_2+1/\alpha_3=3$, which implies that $\alpha_3=\alpha_1%
\alpha_2(3\alpha_1\alpha_2-\alpha_1-\alpha_2)^{-1}$, for arbitrary values of 
$\alpha_1$ and $\alpha_2$. (In particular, we can take simply $%
\alpha_1=\alpha_2=\alpha_3=1$).

Collecting the results we have that our commutative $N=2$ case solution reads%
\begin{equation}  \label{solN2comm}
q(x)=\text{tanh}(2x+c)+ \widehat{i^{i}}\frac{1}{\sqrt{3\alpha_{i}}}\;\text{%
sech}(2x+c)
\end{equation}

As expected, in the present case the structure of the solutions is much
richer than in the complex case, as we have here the parameters $\alpha_{i}$
to play with. Applying the usual parameterization $\Lambda(x)=$tanh$(2x+c)$,
the solution takes the form%
\begin{equation}  \label{solN2commL}
q(x)=\Lambda(x)+\widehat{i^{i}}\frac{1}{\sqrt{3\alpha_{i}}}\;\sqrt {%
1-\Lambda(x)^{2}},
\end{equation}
which looks more suitable for a geometrical analysis.


\subsection{Quaternionic spacetime equivalent solutions}

Let us now evaluate a quaternionic base manifold (noncommutative spacetime
equivalent). This can be done by identifying the spacetime spatial
coordinate $x$ with one of the complex directions of a quaternionic
manifold, namely 
\begin{equation}
x\rightarrow\widehat{i}_{1}X^{1} \quad(i.e. \; x\in SU(2)).
\end{equation}
Therefore, we shall consider scalar fields of the target space depending on
this coordinate, and we must also specialize the operator $\Pi$ to the $%
X^{1} $ coordinate, that is 
\begin{equation}
\Pi\equiv\widehat{i^{0}}\partial_{0}-\widehat{i^{1}}\partial_{1}-\widehat{%
i^{2}}\partial_{2}-\widehat{i^{3}}\partial_{3} \;\;\rightarrow \quad\Pi=-%
\widehat{i^{1}}\partial_{1}
\end{equation}
As a consequence of this choice, the spacetime assumes the structure $%
S_1\otimes O(3)\sim S_1 \otimes SU(2)$, perfectly described by an element $X$
of $\mathbb{H}$, in the representation that we have adopted in this work.

Now, in the present case, due to the non trivial topology of the base
manifold (spacetime equivalent), we solve the first order equations directly
(no trial orbit method). Then, the two cases analyzed before ($N=1$ and $N=2$%
), take the completely different form described below.


\subsubsection{Case $N=1$ (noncommutative)}

Taking $x\rightarrow \widehat{i}_{1}X^{1}$ in the case $N=1$ leaves the
first order equation (\ref{1stord}) with the form 
\begin{eqnarray}
\Pi q=\overline{W^{\prime }\left( q\right) } &=&n-\overline{q} \\
-\widehat{i^{1}}\partial _{_{1}}\left[ q_{0}\!+\!\widehat{i}%
_{_{1}}q_{_{1}}\!+\!\widehat{i}_{_{2}}q_{_{2}}\!+\!\widehat{i}_{_{3}}q_{_{3}}%
\right] &=&n-q_{0}\!+\!\widehat{i^{1}}q_{1}\!+\!\widehat{i^{2}}q_{2}\!+\!%
\widehat{i^{3}}\,q_{3},  \notag
\end{eqnarray}%
which can be put in the form 
\begin{align}
\frac{dq_{0}}{dX_{1}}& =-q_{1} \\
\frac{dq_{1}}{dX_{1}}& =n-q_{0} \\
\frac{dq_{2}}{dX_{1}}& =-q_{3} \\
\frac{dq_{3}}{dX_{1}}& =q_{2}.
\end{align}%
As the equations are not all coupled but in pairs, the resolution of the
system is quite direct, and we obtain the solution 
\begin{eqnarray}  \label{solN1ncomm}
q_{0}(X_{1}) &=&n-e^{C_{+}-C_{-}}\cosh (X_{1}+C_{+}+C_{-}) \\
q_{1}(X_{1}) &=&e^{C_{+}-C_{-}}\sinh (X_{1}+C_{+}+C_{-}) \\
q_{2}(X_{1}) &=&C\cos (X_{1}+\varphi ) \\
q_{3}(X_{1}) &=&C\sin (X_{1}+\varphi )
\end{eqnarray}%
where $C$, $C_{+}$, and $C_{-}$ are integration constants. Now, rewriting
the hyperbolic/trigonometric functions by introducing the algebraic
parameterization $\Lambda_{A}=\tanh(X_{1}+C_{+}+C_{-}),\,\Lambda_{B}=%
\tan(X_{1}+\varphi )$, the quaternion solution assumes a much more
appropriate form for a geometrical analysis 
\begin{equation}
q=n-\frac{e^{C_{+}-C_{-}}}{\sqrt{1-\Lambda _{A}^{2}}}(1+\widehat{i}%
_{1}\Lambda _{A})+\frac{C}{\sqrt{1+\Lambda _{B}^{2}}}\left( \widehat{i}_{2}+%
\widehat{i}_{3}\Lambda _{B}\right)  \label{solN1ncommL}
\end{equation}


\subsubsection{Case $N=2$ (non-commutative)}

In this case the first order equation $\Pi\, q =\overline{W^{\prime}(q)}$
can be split up into its \emph{Sc} and \emph{Vec} parts to give the system 
\begin{align}
\frac{dq_{0}}{dX_{1}} & =-2q_{1}q_{0} \\
\frac{dq_{1}}{dX_{1}} & =n-q_{0}^{2}+q_{1}^{2}+q_{2}^{2}+q_{3}^{2} \\
\frac{dq_{2}}{dX_{1}} & =-2q_{3}q_{0} \\
\frac{dq_{3}}{dX_{1}} & =+2q_{2}q_{0}
\end{align}

The explicit symmetry of this system suggests that the simplest non-trivial
solution is $q_{1}=constant$. With that choice we have 
\begin{align}
n & =q_{0}^{2}-(q_{1}^{2}+q_{2}^{2}+q_{3}^{2}) \\
q_{0}(X_{1}) & =C_{0}\,e^{-2X_{1}q_{1}} \\
q_{2}(X_{1}) & =C_{1}\sin\left(
C_{0}q_{1}^{-1}e^{-2X_{1}q_{1}}+\varphi\right) \\
q_{3}(X_{1}) & =C_{1}\cos\left(
C_{0}q_{1}^{-1}e^{-2X_{1}q_{1}}+\varphi\right)
\end{align}
where the integration constants $C_{0}$ and $C_{1}$ must fulfill the
constraint equation 
\begin{equation}
n^{2}=\left( C_{0}\;e^{-2X_{1}q_{1}}\right) ^{2}-(q_{1}^{2}+C_{1}^{2})
\end{equation}
In particular, we can make the convenient choice 
\begin{equation}
C_{0}=\sqrt{n^{2}+2q_{1}^{2}}\quad\text{and}\quad C_{1}=q_{1}\,,
\end{equation}
which simplifies the solution to the expression 
\begin{eqnarray}
q(X_{1})&=&q_{0}(X_{1})+q_{1}( \widehat{i^{1}}+\widehat{i^{2}}\sin(
q_{0}q_{1}^{-1}+\varphi)  \notag \\
&&+\widehat{i^{3}}\cos( q_{0}q_{1}^{-1}+\varphi) ) \quad
\end{eqnarray}
where 
\begin{equation}
q_{0}(X_{1})=\sqrt{n^{2}+2q_{1}^{2}}\;e^{-2X_{1}q_{1}}.
\end{equation}
This expression can be rewritten in an algebraic form using the
parameterization $\Phi=\tan\left( q_{0}q_{1}^{-1}+\varphi\right) $. We
obtain 
\begin{equation}  \label{solN2noncomm}
q(X_{1})=q_{0}(X_{1})+q_{1}\left[ \widehat{i^{1}}+\frac{1}{\sqrt{1+\Phi^{2}}}%
\left( \widehat{i^{2}}+\widehat{i^{3}}\Phi\right) \right]
\end{equation}

\medskip Another simple solution is obtained by putting $q_{0}\equiv 0$,
which results in 
\begin{equation}
\begin{array}{rcl}
\frac{dq_{1}}{dX_{1}} & = & n+q_{1}^{2}+q_{2}^{2}+q_{3}^{2}\; \\ 
q_{2} & = & C_{2} \\ 
q_{3} & = & C_{3}.%
\end{array}%
\end{equation}
Thus, 
\begin{equation}
q_{1}(X_{1})=\sqrt{n+{C_{2}}^{2}+ {C_{3}}^{2}}\;\tan(X_{1}+C_{1})
\end{equation}

The compactness of the solution is quite evident, even more putting $%
\Phi=\tan(X_{1}+C_{1})$, which gives 
\begin{equation}
q\left( X_{1}\right) =\widehat{i^{1}}\sqrt{n+C_{2}^{2}+C_{3}^{2}}\;\Phi+%
\widehat{i^{2}}C_{2}+\widehat{i^{3}}C_{3}.
\end{equation}


\section{ $GQA$\ and hyper-K\"{a}hler Q-structures}

In this paper the case with torsion will not be considered. Nevertheless, it
is worth noting the particular cases of interest:

\begin{itemize}
\item[i)] $\mathcal{N}=(2,0)$, $D=4$ (or $\mathcal{N}=(4,4)$, $D=2$): the
torsion vanishes, complex structures are annihilated by covariant
derivatives and form the quaternionic algebra (hyper-K\"{a}hler geometry).
In both cases, prepotentials are known in the seminal references \cite%
{ivanov, sokatchev}.

\item[ii)] $\mathcal{N}=(4,0)$, $D=2$: the torsion is a closed 3-form;
complex structures are annihilated by covariant derivatives and form the
quaternionic algebra (hyper-K\"{a}hler with torsion).
\end{itemize}

Then, in the following we will be dealing with the first case, just to show
the consistency of the procedure of finding BPS solutions in the context of
SUGRA.


\subsection{The new metrics: Quink, a Q-Kink analogue?}

The general form of the metrics we are interested in, admitting a
quaternionic structure (hyper-K\"{a}hler or not) is, following the notation
of \cite{TP}, 
\begin{equation}
ds_{new}^{2}=Udq\cdot dq+U^{-1}\left( dq_{0}+\omega \cdot dq\right) ^{2}
\label{metricU}
\end{equation}%
Now, as in \cite{TP}$\omega $ can be considered as an euclidean 3-vector and
it must fulfill the Killing's equation as follows 
\begin{equation}
\mathcal{L}_{X}E_{0}=0,  \label{killingeq}
\end{equation}%
where $E_{0}=U^{-1}(dq_{0}+\omega \cdot dq)\equiv U^{-1}(dq_{0}+\widetilde{%
\omega })$ is the tetrad one-form corresponding to the scalar component of
the line element (\ref{metricU}) and the Killing vector field $X$ is given
i.e. by $\partial _{q_{0}}$. Notice that, from the point of view of the
Cartan's structure equations, relation (\ref{killingeq}) can interpreted as
an integrability condition.

\smallskip As our case does not include torsion (hyper-K\"{a}hler and
Conformal hyper-K\"{a}hler target spaces), clearly, the relation between $U$
and $\widetilde{\omega }\equiv $ $\omega \cdot dq,$ is 
\begin{equation}
Ud\left( U^{-1}\right) =\widetilde{\omega }d\left( \widetilde{\omega }%
^{-1}\right)
\end{equation}

As shown in detail in \cite{tobepub} a connexion (harmmonic map) between the
potential in our $GQL$ and the free sigma model can be explicitly
established, leading to 
\begin{equation}
V = [det(g_{ab})]^{-1},
\end{equation}
where $g_{ab}$ is the metric associated to the K\"{a}hler manifold (target
space) of the free model. Consequently, we can write a simple relation
between our $GQL$ potential and the $U$ factor of the metrics (\ref{metricU}%
), namely 
\begin{equation}
V=\frac{m^{2}}{4}U^{-1}.
\end{equation}
which corroborates the important results from physical arguments given in 
\cite{Sezgin, TP}, now obtained from a purely geometrical approach.


\subsubsection{Metric for the $N=1$ case (commutative)}

Let us consider first the $N=1$ case. The potential is related to the $U$
factor by 
\begin{equation}
V=\frac{1}{2}(n-\overline{q})(n-q)=\frac{m^{2}}{4}U^{-1}.
\end{equation}
Then we can write 
\begin{equation}
U=\det\left( g_{\alpha\beta}\right) =\frac{m^{2}/2}{n^{2}-|q|^{2}}.
\end{equation}
Then, the line element corresponding to the quaternionic solution (\ref%
{solN1comm}) reads 
\begin{equation}
ds^{2}=U^{-1}C_{0}^{\,2}\,e^{-2x}d^{2}x+UC_{i}^{\,2}\,e^{2x}\sigma^{i}%
\otimes\sigma^{i},
\end{equation}


\subsubsection{Metric for the $N=2$ case (commutative)}

Let us now consider $N=2$ case. For the potential we can write%
\begin{equation}
V=\frac{1}{2}(n-\overline{q}^{2})(n-q^{2})=\frac{m^{2}}{4}U^{-1}.
\end{equation}
Therefore, we have the relation 
\begin{equation}
U=\det\left( g_{\alpha\beta}\right) =\frac{m^2/2}{n^{2}-|q|^{4}},
\end{equation}
Similarly to the $N=1$ case, we can write the line element corresponding to
the quaternionic solution (\ref{solN2comm}) calculated above, 
\begin{equation}
ds^{2}=4\, \text{sech}^{4}(2x+c)\left[U^{-1} dx^{2}+U\frac{%
\sigma^{i}\otimes\sigma^{i}}{3\alpha_{i}}\sinh^{2}(2x+c)\right]
\end{equation}
The analysis of this expression is extremely simplified introducing the
relation $\Lambda\equiv \tanh(2x+c)$ as in (\ref{solN2commL}). Such
definition transforms the hyperbolic/trigonometric expressions in algebraic
ones leading to the line element%
\begin{equation}
ds^{2}=4\left(1-\Lambda^{2}\right) ^{2}\left[ U^{-1}dx^{2}+U\frac{%
\sigma^{i}\otimes\sigma^{i}}{3\alpha_{i}} \frac{\Lambda^{2}}{%
\left(1-\Lambda^{2}\right)}\right]
\end{equation}


\subsection{Generalization of the Berger's sphere and comparison with other
solutions}

Let us now analyze the 3-dimensional (compact) part of the metrics obtained
above, $ds_{3}^{2}=Udq\cdot dq$. In order to make explicit the $S_{1}\otimes
S_{2}$ structure, we introduce the usual left angle-variables representative
forms of the compact submanifold, and some constant coefficients $\overset{%
\circ}{q}_{i}$. Two main consequences immediately arise:

\smallskip i) If $\overset{\circ}{q}_{1}=\overset{\circ}{q}_{2}=\overset{%
\circ}{q}_{3}$, that is, if the compact part of the metric takes the form 
\begin{eqnarray}
{ds_{3}^{2}}_{(N=1)}\!\!&=&\!\!U C_{i}^{2}\,e^{2x}\overset{\circ}{q_{1}}^{2} %
\left[d\theta^{2}+ \sin^{2}\theta d\varphi^{2} \right. \\
&&\quad\left. +( d\psi+\cos\theta d\varphi)^{2}\right]  \notag \\
{ds_{3}^{2}}_{(N=2)}\!\!&=&\!\!4U (1-\Lambda^{2}) \Lambda^{2}\overset{\circ}{%
q_{1}}^{2} \left[d\theta^{2} +\sin^{2}\theta d\varphi^{2}\right. \\
&&\quad\left.+(d\psi+\cos\theta d\varphi)^{2}\right]  \notag
\end{eqnarray}
then the solution is a generalization similar to the Abraham-Townsend
(Q-Kink) solution \cite{AT}, and to the multicenter Gibbons-Hawking \cite{GH}
solution, which are hyper-K\"{a}hler or conformally hyper-K\"{a}hler target
manifolds.

\smallskip ii) If $\overset{\circ}{q}_{1}=\overset{\circ}{q}_{2}\neq\overset{%
\circ}{q}_{3}$, then the metric corresponds to a \textit{generalization of
the Berger's sphere} (deformation of the $O(3)\approx SU\left( 2\right)$
structure), which is a smooth deviation from the pure hyper-K\"{a}hler
condition, 
\begin{eqnarray}
{ds_{3}^{2}}_{(N=1)}\!\! &=&\!\!U C_{i}^{2}\,e^{2x}\overset{\circ}{q_{1}}\!%
\left[\overset{\circ}{q_{1}}^{2}(d\theta^{2}+ \sin^{2}\theta d\varphi^{2})
\right. \\
&&\quad\left.+\overset{\circ}{q_{3}}^{2}( d\psi+\cos\theta d\varphi)^{2}%
\right]  \notag \\
{ds_{3}^{2}}_{(N=2)}\!\! &=&\!\!4U(1-\Lambda^{2}) \Lambda^{2}\!\left[\overset%
{\circ}{q_{1}}^{2}(d\theta^{2}+ \sin^{2}\theta d\varphi^{2})\right. \\
&&\quad\left. +\overset{\circ}{q_{3}}^{2}(d\psi+\cos\theta d\varphi)^{2}%
\right]  \notag
\end{eqnarray}

We can now establish a comparison between our soultions and the well known
results from other works \cite{AT, GH} -- see Table \ref{tab:sol}.

Finally, note that in the cases of the commutative spacetime equivalent, the
obtained metrics are generalizations of Majumdar-Papapetrou solutions.%


\begin{center}
\begin{table*}[h!t]
\begin{tabular}{|c|c|c|c|}
\hline
\multicolumn{2}{|c|}{\bfseries AT} & \multicolumn{2}{|c|}{{\bfseries GQSM}
(commutative)} \\ \hline
{\bfseries \quad One center \quad} & {\bfseries Two centers} & {$\mathbf{N}%
=1 $} & {$\mathbf{N}=2$} \\ \hline
$\quad U=2m\frac{1}{|q-q_{0}|}\quad$ & $\quad U=2m\left[\frac{1}{|q-q_{0}|}+%
\frac{1}{|q+q_{0}|}\right]\quad$ & $U=\frac{m^{2}/2}{\left(n^{2}-|q|^{2}%
\right)}$ & $U=\frac{m^{2}/2}{\left( n^{2}-|q|^{4}\right)}$ \\ \hline
\multicolumn{2}{|c|}{$%
\begin{array}{c}
q_{0}(x)=const., \\ 
\text{(angular character: compact)}%
\end{array}%
$} & $\quad q_{0}(x)=n-C_{0}\,e^{-x}\quad$ & $\quad q_{0}(x)=\tanh\left(%
\frac{m\left\vert \eta\right\vert }{8\mu}(2x+c)\right) \quad $ \\ \hline
\multicolumn{2}{|c|}{$%
\begin{array}{c}
{q}_{i}(x)=\Phi\tanh\left( \frac{m\left\vert \eta\right\vert }{8\mu}%
(x-x_{0})\right), \\ 
\Phi=const.%
\end{array}%
$} & $q_{i}(x)=C_{i}\,e^{x}$ & $%
\begin{array}{c}
q_{i}(x)=\overset{\circ}{q}_i \,\text{sech}\left(\frac{m|\eta|}{8\mu}%
(2x+c)\right), \\ 
\overset{\circ}{q}_{i}=1/\sqrt{3\alpha_i} \quad (\alpha_{i}\geqq 0\; \forall
i) \\ 
\quad\text{and}\quad 1/\alpha_1+1/\alpha_2+1/\alpha_3=3%
\end{array}%
$ \\ \hline
\end{tabular}%
\caption{Comparison with solutions from Abraham-Townsend (AT) model.}
\label{tab:sol}
\end{table*}
\end{center}


\vspace{-20pt}

\section{More examples: Generalization of standard metrics factorization}

It is well known, in the context of SUGRA, the importance of the metrics
that can be factorized as product of lower dimensional ones. The main reason
is the claim that the appearing of these type of metrics (in particular the
Bertotti-Robinson's ($BR$) and generalizations) in supergravity theories,
indicates that the theory is fully renormalizable. The proof of the
non-renormalization theorem for the $BR$ background was almost trivial due
to conformal flatness of this type of metrics, and because the Maxwell field
is constant. These properties are not present in the general case of metrics
admitting super-covariantly constant spinors. In General Relativity these
solutions are known as the conformal-stationary class of Einstein-Maxwell
fields, with conformally flat 3-dimensional spaces. Some generalizations of
this class of metrics have been found by Neugebauer, Perjes, Israel and
Wilson \cite{NPIW}: the flat space Laplacian in x. However, the analysis of
these subjects is out of the scope of this letter 
, and will be discussed elsewhere.

The product type metrics are 4-dimensional but composed by two 2-dimensional
ones, in general, of K\"{a}hler type. In particular, we have found two
metrics showing this structure in general form , as described below.


\subsection{Case $N=1$ (noncommutative)}

For this case we have obtained a metric solution of the form 
\begin{eqnarray}
ds^{2} &=&\frac{e^{2\left( C_{+}-C_{-}\right) }}{1-\Lambda _{A}^{2}}\left(
U^{-1}\Lambda _{A}^{2}dx_{1}^{2}+U\sigma ^{1}\otimes \sigma ^{1}\right) 
\notag \\
&&+\frac{C_{1}^{2}}{1+\Lambda _{B}^{2}}U\left( \Lambda _{B}^{2}\sigma
^{2}\otimes \sigma ^{2}+\sigma ^{3}\otimes \sigma ^{3}\right) .
\end{eqnarray}%
This metric `product' is the result of the geometrical structure of the
quaternion solution (\ref{solN1ncomm}). Precisely, if we make the choice $%
n=0 $ , $C_{+}=C_{-}$ and $i2C_{+}=\varphi $, the quaternion solution (\ref%
{solN1ncommL}) can be written as the result of a product of the form $%
\mathbb{C}\otimes \mathbb{C}\rightarrow \mathbb{H}$, as follows 
\begin{equation}
q=\left[ \mathbb{I}_{2}+f\sigma ^{2}\right] z\,,
\end{equation}%
where the complex number $z\in \mathbb{C}\subset \mathbb{H}$ is defined as 
\begin{equation}
z\equiv \cosh \left( x_{1}+C_{+}+C_{-}\right) +\sinh \left(
x_{1}+C_{+}+C_{-}\right) \sigma ^{1}\,,
\end{equation}%
and the mapping $f$ over the $\mathbb{C}$ field is given by 
\begin{equation}
f:\mathbb{C}\left( x_{1}+C_{+}+C_{-}\right)\;\; \rightarrow \;\; \mathbb{C}%
\left( i\left( x_{1}+C_{+}+C_{-}\right) \right) .
\end{equation}%
This is the reason why in this case the metric can be interpreted as product
of two complex K\"{a}hler metrics with the structure $\mathbb{C}\otimes 
\mathbb{C}\rightarrow \mathbb{H}$


\subsection{Case $N=2$ (noncommutative)}

For this case we have obtained a metric solution of the form 
\begin{eqnarray}
ds^{2} &=&4q_{0}^{2}q_{1}^{2}\left( U^{-1}dx_{1}^{2}+\frac{U}{q_{0}^{2}}%
\sigma ^{1}\otimes \sigma ^{1}\right) \\
&&\quad +\frac{U}{1+\Phi ^{2}}\left( \sigma ^{2}\otimes \sigma ^{2}+\Phi
^{2}\sigma ^{3}\otimes \sigma ^{3}\right) .  \notag
\end{eqnarray}%
This metric `product' is the result of the geometrical structure of the
quaternionic solution (\ref{solN2noncomm}). Notice that now, in a sharp
contrast with the previous case, the solution cannot be written directly as
the result of a product of the form $\mathbb{C}\otimes \mathbb{C}\rightarrow 
\mathbb{H}$, or as the action of an ideal over a complex field $\in \mathbb{H%
}$.


\section{Concluding remarks}

In the present letter we have proposed a new method for finding BPS
solutions in the context of SUGRA. This method, in sharp contrast with the
methods of the calibrations \cite{papad1}, or the monopole method introduced
in \cite{papad2, CTV}, allows to find suitable Quaternionic and hyper-K\"{a}%
hler geometries with the required properties appearing in the bosonic sector
of supergravity theories based in the susy extensions of the nonlinear sigma
models. The solutions found have the particularity of being BPS and are not
just generalizations of the well known solutions, but new distinct ones.

While the focus of the present work was to present the method, that is, the
geometrical link between genuine BPS\ quaternionic solutions of the $GQL$\
and the target space metric of the bosonic sector of the SUGRA theory, it is
important to note the clear existence of a relation between our solutions
and the Bianchi $IX$ generalizations of hyper-K\"{a}hler metrics with
Taub-NUT structures, and also the possible close connexion of our method
with the (unconstrained) Harmonic Superspace Formalism \cite{ivanovSS}. The
proper analisys of these points requires the evaluation of the full version
of the supergravity theory (bosonic and fermionic sectors), and will be
addressed in future work.


\section{Acknowledgments}

We would like to thank CNPq and PROCAD/CAPES for partial financial support.


\section{Appendix. Energy, left regular $W(q)$ and BPS conditions}

\label{Appndx} Similarly to the complex case of Ref. \cite{complex}, we can
rewrite the energy in a convenient fashion, in order to make explicit the
relation between the BPS conditions and the corresponding gradient and
potential terms of the Hamiltonian. Namely, we have 
\begin{eqnarray}
E&=&\tfrac{1}{2}\int dx\biggl[ \biggl( \frac{dq_{0}}{dx}+W_{q_{i}}\biggr) 
\overline{\biggl( \frac{dq_{0}}{dx}+W_{q_{i}}}\biggr)  \notag \\
&&+\biggl( \frac{dq_{i}}{dx}+W_{q_{0}}\biggr) \overline{\biggl( \frac{dq_{i}%
}{dx}+W_{q_{0}}\biggr)}\biggr]  \notag \\
&&\quad -\int dx\ Sc\biggl( \frac{dq_{0}}{dx}\overline{W_{q_{i}}}+\frac{%
dq_{i}}{dx}\overline{W_{q_{0}}}\biggr)  \label{energy}
\end{eqnarray}

Here, consistently with the 1-dimensional spatial coordinate, we retain the
scalar (commuting) part of the $\Pi$ operator and of the quaternionic
position $X$ introducing the usual (commutative) $x$ coordinate: $%
\Pi_{0}\rightarrow\frac{d}{dx},X_{0}\rightarrow x$.

For quaternionic field solutions obeying 
\begin{eqnarray}
\frac{dq_{0}}{dx}&=&-W_{q_{i}}, \\
\frac{dq_{i}}{dx}&=&-W_{q_{0}},
\end{eqnarray}
expression (\ref{energy}) is minimized to the Bogomol'nyi bound, and the
energy is given by the superpotential 
\begin{equation}
E_{BPS}=\int dX \left( |W_{q_{i}}|^{2} + |W_{q_{0}}|^{2}\right)
\label{BPSbound}
\end{equation}
Analogously to the complex case, where Cauchy-Riemann conditions arise, the
above equations solve the quaternionic equation of motion if we impose the
Fueter-harmonic condition on $W(q_{0},q_{i})$, that is 
\begin{eqnarray}
0&=&\square\overline{W\left(q_{0},q_{i}\right)}=\!\partial_{q_{0}}^{2}%
\overline{W}+\partial_{q_{1}}^{2}\overline{W}+\partial_{q_{2}}^{2}\overline{W%
}+\partial_{q_{3}}^{2}\overline{W}  \notag \\
&=&\!\square W\left(q_{0},q_{i}\right) -2{\Pi}_{q}\left[Sc\left(W\left(
q_{0},q_{i}\right) \right) \right].
\end{eqnarray}
That is, the Fueter-harmonic condition implies the (left) holomorphy of $W$%
\begin{eqnarray}  \label{lefthol}
0&=&\Pi_{q}\overline{W\left( q_{0},q_{i}\right) } \\
&=&\overline{\Pi }_{q}W\left(q_{0},q_{i}\right) -2 {\Pi}_{q} \left[ Sc\left(
W\left( q_{0},q_{i}\right)\right) \right].  \notag
\end{eqnarray}
Therefore, we can generalize previous results concerning the complex fields 
\cite{complex}, to a quaternionic function $\mathcal{W=}\widetilde{W}+W$,
such that the Cauchy-Fueter left regularity is satisfied%
\begin{equation}
\widetilde{W}_{q_{0}}=-W_{q_{i}},\qquad \widetilde{W}_{q_{i}}=-W_{q_{0}}
\end{equation}
It is worth mentioning here that all the above expressions involving
analytical properties of the functions in the quaternionic field, reduce to
their corresponding analogous expressions in the complex field case, when we
retain only two of the quaternionic variables, namely%
\begin{equation}
W\left( q_{0},q_{i}\right) \;\rightarrow\; W\left(
q_{0},q_{1}\right)=W_{0}+iW_{1}
\end{equation}
In this case, equation (\ref{lefthol}) reduces to the Cauchy-Riemmann
conditions 
\begin{eqnarray}
\left(\partial_{q_{0}}+i\partial_{q_{1}}\right)\left(W_{0}+iW_{1}\right)-2i%
\partial_{q_{1}}W_{0}&=&0  \notag \\
\left( \partial_{q_{0}}W_{0}-\partial_{q_{1}}W_{1}\right) +i\left(
\partial_{q_{0}}W_{1}-\partial_{q_{1}}W_{0}\right) & =&0 \\
\notag
\end{eqnarray}%
%
%
%
%
%

Taking into account the above statements, the energy can be put in a more
general fashion%
\begin{eqnarray}  \label{energyW}
E&=&\tfrac{1}{2}\int dX\left[\left(\widetilde{W}_{q_{0}}+W_{q_{i}}\right) 
\overline{\left(\widetilde{W}_{q_{0}}+W_{q_{i}}\right)}\quad\right.  \notag
\\
&&\left.+ \left(\widetilde{W}_{q_{i}}+W_{q_{0}}\right)\overline{\left(%
\widetilde{W}_{q_{i}}+W_{q_{0}}\right)}\right] \\
&&\quad-\int dX\ Sc\left(\widetilde{W}_{q_{0}}\overline{W_{q_{i}}}+%
\widetilde{W}_{q_{i}}\overline{W_{q_{0}}}\right),  \notag
\end{eqnarray}
generalizing equation (\ref{energy}).


\end{document}